\documentclass[12pt]{article} %hier wegkommentieren

\usepackage[ngerman,english]{babel}
\useshorthands{"}
\addto\extrasenglish{\languageshorthands{ngerman}}
\usepackage[utf8]{inputenc}
\usepackage[T1]{fontenc}
\usepackage{amsmath}
\usepackage{graphicx}
\usepackage{amssymb}
\usepackage{hyperref}
\usepackage{mathrsfs}
\usepackage{color}

\renewcommand{\phi}{\varphi}
\renewcommand{\theta}{\vartheta}

\title{{\Large\bf Entropy Counting from Schwarzschild/CFT and Soft Hair}}
\author{{\bf Artem Averin$^{\textrm{a,b}}$\footnote{artem.averin@campus.lmu.de}}}

\begin{document}

\maketitle

%\vskip-2.5cm
\centerline{\it $^{\textrm{a}}$ Arnold--Sommerfeld--Center for Theoretical Physics,}
\centerline{\it Ludwig--Maximilians--Universit\"at, 80333 M\"unchen, Germany}
\medskip
\centerline{\it $^{\textrm{b}}$ Max--Planck--Institut f\"ur Physik,
Werner--Heisenberg--Institut,}
\centerline{\it 80805 M\"unchen, Germany}
%\medskip
%\centerline{\it $^{\textrm{c}}$ Center for Cosmology and Particle Physics,
%Department of Physics, New York University}
%\centerline{\it 4 Washington Place, New York, NY 10003, USA}
%\medskip
%\centerline{\it $^{\textrm{d}}$ Instituto de F\'{\i}sica Te\'orica UAM-CSIC, C-XVI,
%Universidad Aut\'onoma de Madrid,}
%\centerline{\it Cantoblanco, 28049 Madrid, Spain}

\vskip1cm
\begin{abstract}
{{
We revisit Carlip's approach to entropy counting. This analysis reemerged in a recently obtained Schwarzschild/CFT-correspondence as Sugawara-construction of a 2D stress-tensor. Here, for the example of a Schwarzschild black hole, we show how to single out diffeomorphisms forming in contrast to Carlip's analysis the full 2D local conformal algebra. We provide arguments, why their Hamiltonian generators are expected to be the symmetry generators of a possible conformal field theory describing the part of phase space responsible for black hole microstates. Then, we can infer central charges and temperatures of this CFT by inspecting the algebra of these Hamiltonian generators. Using this data in the Cardy-formula, precise agreement with the Bekenstein-Hawking entropy is found. Alternatively, we obtain the same CFT temperatures by thermodynamic considerations. Altogether, this suggests that the Hamiltonian generators need no corrections through possible non-canonical counterterms. We comment on the related recent work by Haco, Hawking, Perry, Strominger.  
}}
\end{abstract}

\begin{flushright}
%\vskipcm
%{\small  MPP--2016--3},
%{\small LMU--ASC 05/16}
\end{flushright}

\newpage

\setcounter{tocdepth}{2}
\tableofcontents
\break

\section{Introduction}
\label{Kapitel 1}

It is one of the main problems in quantum gravity to explain the microcanonical origin of the Bekenstein-Hawking entropy $S=\frac{A}{4 \hbar G}$ of black holes \cite{Bekenstein:1973ur,Hawking,Hawking:1974rv}. In the classical $\hbar \to 0$ limit the entropy becomes infinite. Therefore, the Hamiltonian phase space of pure Einstein gravity has to contain infinitely many points corresponding to the microstates of a black hole for fixed mass and angular momentum parameter. On the other hand, the black hole uniqueness theorems \cite{Heusler} tell that asymptotically flat and stationary solutions of Einstein's field equations are given by the Kerr-family up to diffeomorphisms. This can create the impression of an arising paradox: Due to the uniqueness theorems, it may naively seem that there is no place in Hamiltonian phase space that can accomodate the infinitely many microstates as required by the classically infinite entropy. Therefore, one can ask: How to reconcile the black hole uniqueness theorems with the classically infinite Bekenstein-Hawking entropy?

Since the uniqueness theorems single out the Kerr solutions up to diffeomorphisms, they already themselves suggest a possible solution. It may be that some of the diffeomorphisms are physical, i.e. shifts in phase space rather than gauge redundancies. This phenomenon is known to happen in gauge theories typically when the gauge parameters are non-vanishing in some asymptotic region. Such asymptotic symmetries could in gravity then be responsible for microstates of a Kerr black hole. 

Indeed, the study of asymptotic symmetries in 3D gravity \cite{Brown:1986nw} brought some success in understanding the BTZ black hole. It is found that the asymptotic symmetry algebra contains the 2D local conformal algebra. Conformal field theory techniques can then be used to count the state degeneracy \cite{Strominger:1997eq} and agreement with the Bekenstein-Hawking entropy is found. Carlip raised the idea \cite{Carlip:1998wz,Carlip:1999cy,Carlip:2011ax,Carlip:2011vr} to mimic this in the higher-dimensional case. Although it is not clear which gauge transformations are responsible for microstates, Carlip was able to single out a Witt-algebra of diffeomorphisms in the presence of a black hole event horizon. The Hamiltonian generators of these diffeomorphisms are then candidates for the generators of a possible conformal symmetry that may govern the part of phase space responsible for black hole microstates. Hamiltonian methods can then be used to study the conformal algebra of the diffeomorphism generators and CFT techniques then to count the state degeneracy. Indeed, agreement with the expected Bekenstein-Hawking entropy is found. 

Although Carlip's approach is universal, it tells nothing about what the possible underlying CFT describing the relevant part in phase space really is. To understand this part in phase space was always one of the main motivations in the study of asymptotic symmetries (see for instance \cite{Barnich:2010eb} and references therein). The idea recently gained new interest as the proposal of ``soft black hole hair'' \cite{Hawking:2016msc}.\footnote{The most simplest choice, the $\mathfrak{bms}_4$-supertranslations contained in the asymptotic symmetry algebra of spacetimes that are asymptotically flat at null infinity \cite{Barnich:2010eb}, does not work. It was already explained in \cite{Averin:2016ybl} that they can not be responsible for microstates. Instead, it was proposed there that the presence of an event horizon enhances the asymptotic symmetry algebra and it is the \emph{enhancement} that is responsible for microstates and entropy counting \cite{Averin:2016hhm}. This provides the resolution to the criticism on the soft hair proposal stated later correctly in \cite{Mirbabayi:2016axw1,Mirbabayi:2016axw2,Mirbabayi:2016axw3,Mirbabayi:2016axw4}.} To analyze the structure of Hamiltonian phase space in the vicinity of a black hole state is nevertheless still a necessary and open problem.

To improve the situation is the overall goal of our investigations and in \cite{Averin:2018owq} we described, how the Hamiltonian phase space can be analyzed in a systematic way. For a Schwarzschild black hole and assuming in a sense the application of the simplest possible scenario, we proposed a concrete candidate of a dual theory describing the part of phase space responsible for microstates. This theory was given in terms of its observables and their Poisson-bracket algebra. If conformally invariant, as it is expected from several directions, Carlip's approach to entropy counting reemerges at this point as a Sugawara-construction of the conformal generators of this Schwarzschild/CFT correspondence out of its observables. Accordingly, one has to find a suited choice of Witt-algebra of diffeomorphisms. Carlip presented a general construction of such an algebra in the presence of a black hole. 

Here, we want to revisit Carlip's approach. In Carlip's construction only one copy of a Witt-algebra of diffeomorphisms and associated Hamiltonian generators are found. The two-dimensional conformal algebra consists however of two commuting copies. Since there seems to be no reason, why black holes should be described by chiral CFTs, it is natural to seek for diffeomorphisms building two Witt-algebra copies. Is such a choice possible and does it lead to something maybe even more appropriate? 

That such a choice is possible follows directly from \cite{Averin:2018owq}. There, we have provided a $Vir \oplus \overline{Vir}$-algebra of diffeomorphisms. Using this choice, the entropy counting procedure in the context of the proposed Schwarzschild/CFT-correspondence was discussed. The main idea in the construction of this $Vir \oplus \overline{Vir}$-diffeomorphisms was to still follow Carlip's construction \cite{Carlip:1999cy} closely. But whereas Carlip singles out $Vir$-diffeomorphisms in the presence of a local Killing horizon, we insist in treating both the future and past event horizon of the black hole on the same level. This then leads to two copies of Witt-algebra diffeomorphisms. 

Our purpose here is to report further on our investigations whether this choice of $Vir \oplus \overline{Vir}$-diffeomorphisms is a proper one. This question is in principle independent of the issues discussed in \cite{Averin:2018owq}. Our $Vir \oplus \overline{Vir}$-diffeomorphisms are of interest as they provide a novel choice of diffeomorphisms to be used in Carlip's approach. Inspecting the algebra of Hamiltonian generators, we will in this work infer central charges and Virasoro zero-modes (or equivalently CFT temperatures) that reproduce via Cardy-formula the expected Bekenstein-Hawking entropy. There appears no need to correct the canonically-derived Hamiltonian generators by any counterterms. Furthermore, the derived CFT temperatures are in agreement with the temperatures obtained directly from the $Vir \oplus \overline{Vir}$-vectorfields by thermodynamic considerations. 

The recent work by Haco, Hawking, Perry, Strominger \cite{Haco:2018ske} also follows Carlip's approach to black hole entropy counting. There, an alternative choice of $Vir \oplus \overline{Vir}$-diffeomorphisms for the case of a Kerr black hole is proposed. We comment on that choice throughout our investigations. 

The paper is organized as follows. In chapter \ref{Kapitel 2} we briefly review Carlip's approach to black hole entropy counting. To inspect the algebra of the Hamiltonian generators of diffeomorphisms, we derive the relevant formulas. Especially, we explain how the CFT data needed in the Cardy-formula is derived once a choice of $Vir \oplus \overline{Vir}$-diffeomorphisms has been made. In chapter \ref{Kapitel 3}, for the example of a Schwarzschild black hole, we explain how to single out a ``preferred'' $Vir \oplus \overline{Vir}$-algebra of spacetime diffeomorphisms. Alternatively to the considerations in chapter \ref{Kapitel 2}, we fix the associated CFT temperatures by some thermodynamic considerations. The role of the Casimir-operators of the global conformal algebra for scattering off a black hole is briefly discussed. In chapter \ref{Kapitel 4}, we use the derived $Vir \oplus \overline{Vir}$-diffeomorphisms in the framework of chapter \ref{Kapitel 2} to infer the relevant CFT data for entropy counting. We find agreement with the Bekenstein-Hawking entropy. In chapter \ref{Kapitel 5}, we connect these findings with \cite{Averin:2018owq}. 

In the following, we use units in which we set the speed of light to $1$ but we keep Newton's constant $G$ and Planck's constant $\hbar$ explicit. Latin letters $a, b, \ldots = 0, \ldots, 3$ denote spacetime indices. 

\section{General Argument and Realization}
\label{Kapitel 2}

\subsection{General Argument}
\label{Kapitel 2.1}

Here, we give a brief review of Carlip's approach to explain the statistical mechanical origin and counting of the black hole entropy especially in dimensions higher than $3.$ For a more detailed discussion and references, we refer to the original papers \cite{Carlip:1998wz,Carlip:1999cy,Carlip:2011ax,Carlip:2011vr}. The interpretation of this approach in light of a recently proposed Schwarzschild/CFT-correspondence \cite{Averin:2018owq} was already given in that reference and will also be discussed in chapter \ref{Kapitel 5}.

Consider an arbitrary diffeomorphism-invariant theory of gravity given by some action, which possibly can contain black hole solutions. For a diffeomorphism given by some vectorfield $\xi$ over the spacetime manifold, we denote by $H_\xi$ the associated Hamiltonian generator. The Hamiltonian generator $H_\xi$ - if it exists - is a function over the phase space $\Gamma$ of the theory and implements the diffeomorphism $\xi.$ The generator $H_\xi$ is given as the sum of a bulk integral over suited gauge constraints and a suited boundary integral. On-shell $H_\xi$ is therefore given by a boundary integral. If $H_\xi$ is non-constant over phase space, the diffeomorphism $\xi$ constitutes a physical excitation, i.e. $H_\xi$ implements a shift in the Hamiltonian phase space, otherwise a gauge redundancy. 

The algebra of the Hamiltonian generators $H_\xi$ with respect to the Poisson-bracket forms on-shell a representation of the algebra of the associated diffeomorphisms with respect to the ordinary Lie-bracket of vectorfields over the spacetime manifold up to central extensions. That is, for spacetime vectorfields $\xi_1, \xi_2$ we have on-shell the relation 

\begin{equation}
\left\{ H_{\xi_1}, H_{\xi_2} \right\} = H_{\left[ \xi_1, \xi_2 \right]} + K_{\xi_1,\xi_2}
\label{1}
\end{equation}

where $K_{\xi_1,\xi_2}$ are constant c-numbers. 

So far, we have reviewed general statements of the Hamiltonian mechanics for gravity theories. What happens if we have a black hole solution? In this case the idea is to treat the event horizon as a boundary of the spacetime manifold. The presence of such a boundary can render some diffeomorphisms $H_\xi$ from would-be gauge redundancies to physical excitations which could be important for the statistical mechanics of the black hole. The presence of the boundary can furthermore give rise to non-vanishing central extensions in the algebra \eqref{1} of the aforementioned generators $H_\xi.$ Carlip's observation was that for black hole event horizons there are ``natural'' ways to find diffeomorphisms $\xi_n$ $(n \in \mathbb{Z})$ which form a Witt-algebra

\begin{equation}
\left[ \xi_m, \xi_n \right] = -i(m-n) \xi_{m+n}.
\label{2}
\end{equation}

The subalgebra of \eqref{1} of the associated generators $H_{\xi_n}$ then forms a Virasoro-algebra. 

Virasoro-algebras constitute the symmetry algebras of two dimensional conformal field theories. The assumption then is, that there is a 2D CFT which describes the part of the phase space that is responsible for black hole microstates and whose conformal generators are provided by the $H_{\xi_n}.$ At this stage, it is of course not clear \emph{whether} such a theory exists or \emph{what} this theory is. However, accepting this assumption one has fortunately the luxury that a lot of information about a given 2D CFT can be gained from its Virasoro-algebra - for our black hole case this would then be the algebra of the generators $H_{\xi_n}.$ 

For instance, the degeneracy of states in a 2D CFT is (often) fixed through the Cardy formula by the central charge which is read directly from the Virasoro-algebra. Therefore, it is tempting to perform the following sort of consistency check of the aforementioned assumption. One can determine the central charge of the Virasoro-algebra formed by the $H_{\xi_n}$ by calculating the extensions $K_{\xi_m, \xi_n}$ in \eqref{1}. The associated degeneracy of states of the would-be CFT is then compared with the black hole's Bekenstein-Hawking entropy. Carlip's result was that both of them agree. 

As already mentioned, there are still several remaining open questions. For example, why is it appropriate to treat the event horizon as a boundary in the evaluation of the boundary integrals that determine the extensions $K_{\xi_m, \xi_n}?$ To put it differently, this can be phrased as \emph{what} the CFT governing the black hole microstates in phase space is and \emph{how} to obtain it. To answer these questions is the overlying goal of our investigations and we will briefly come back to these issues in chapter \ref{Kapitel 5} where we emphasize the connection with previous work. 

However, our point in this work is to revisit Carlip's approach to entropy counting in several directions. Why is such a revision necessary?

The symmetry algebra of a 2D CFT contains \emph{two} mutually commuting copies of Virasoro-algebras. The associated chiral and anti-chiral central charges are usually equal (CFTs in curved background with different chiral and anti-chiral central charges are even known to be inconsistent \cite{Qualls:2015qjb}). In Carlip's approach \cite{Carlip:1999cy} instead, the entire contribution to black hole entropy comes from a chiral half of a would-be CFT. 

In this work, we therefore want to present a way to construct diffeomorphisms $\xi_n$ and $\overline{\xi}_n$ $(n \in \mathbb{Z})$ satisfying two copies of the Witt-algebra 

\begin{equation} \label{3}
\begin{split}
\left[ \xi_m, \xi_n \right] &= -i(m-n)\xi_{m+n} \\
\left[ \overline{\xi}_m, \overline{\xi}_n \right] &= -i(m-n)\overline{\xi}_{m+n} \\
\left[ \xi_m, \overline{\xi}_n \right] &= 0.
\end{split}
\end{equation}

The choice of diffeomorphisms should be such that the associated Hamiltonian generators $H_{\xi_n}$ and $H_{\overline{\xi}_n}$ are reasonable candidates for the symmetry algebra of a possible CFT governing the statistical mechanics of the black hole under consideration.

\subsection{Realization}
\label{Kapitel 2.2}

In order to be as simple and as concrete as possible, we consider the case of a Schwarzschild black hole in pure Einstein gravity. We denote spacetime coordinates by $x^a = (x^0,x^1,x^A)$ with angular coordinates indexed by $A,B,\ldots = 2, 3.$ In ingoing Eddington-Finkelstein coordinates $(v,r,x^A) = (v,r,\theta,\phi)$ the Schwarzschild-metric reads

\begin{equation} \label{4}
\begin{split}
ds^2 &= g_{ab} dx^a dx^b \\
&= -\left(1-\frac{r_S}{r}\right)dv^2+2dvdr+r^2\left(d\theta^2+\sin^2 \theta d\phi^2\right) \\
&=-\left(1-\frac{r_S}{r}\right)dv^2+2dvdr+r^2\gamma_{AB}dx^Adx^B.
\end{split}
\end{equation}

They are related to the ordinary Schwarzschild-coordinates $(t,r,\theta,\phi)$ by

\begin{equation}
v = t+r^*,
\label{5}
\end{equation}

where the tortoise coordinate is given by

\begin{equation}
r^* = r + r_S \ln \left|\frac{r}{r_S}-1\right|.
\label{6}
\end{equation}

We work here in Eddington-Finkelstein coordinates because then the metric \eqref{4} satisfies (advanced) Bondi-gauge just to be compatible with our previous conventions in \cite{Averin:2018owq}. However, for what follows this choice is arbitrary and we could work in any coordinates that cover the event horizon. $\gamma_{AB}$ denotes the metric on the unit $2$-sphere. In what follows, we consider the metric \eqref{4} as a fixed reference point $g_{ab} \in \Gamma$ in Hamiltonian phase space describing a Schwarzschild black hole with mass parameter $\frac{r_S}{2G}.$ Our task is then to look at the behavior of Hamiltonian generators of suited diffeomorphisms in the vicinity of this point. 

To study the behavior of these generators, we have to refer to Hamiltonian mechanics. In order to do so, several approaches exist. One way would be to use the direct Hamiltonian approach to general relativity as it was done in \cite{Carlip:1998wz}. However, we will use the covariant phase space formalism \cite{Lee:1990nz,Wald:1999wa} which is manifestly covariant as used in \cite{Carlip:1999cy}. The formalism is for example reviewed in \cite{Seraj:2016cym,Hawking:2016sgy,Compere:2018aar} and we will use some formulas collected there.

In the covariant phase space formalism the Hamiltonian phase space $\Gamma$ is given by the solution space of the theory under consideration (i.e. the set of all field configurations satisfying the equations of motion). In principle, one has to divide out symplectic zero-modes by appropriately fixing the gauge but in our present context this step is not relevant. For a diffeomorphism $\xi$ the infinitesimal change of the associated Hamiltonian generator between points $g_{ab}+h_{ab} \in \Gamma$ and $g_{ab} \in \Gamma$ in phase space is determined by

\begin{equation}
\delta H_\xi \left[h_{ab}; g_{ab} \right] = -\frac{1}{16 \pi G} \oint_{\partial \Sigma}{*F}.
\label{7}
\end{equation}

Here, $\Sigma$ is a Cauchy-surface in the spacetime manifold and the $2$-form $F_{ab}$ is well-known \cite{Seraj:2016cym,Hawking:2016sgy,Compere:2018aar}

\begin{equation} \label{8}
\begin{split}
F_{ab} &= \frac{1}{2}\left(\nabla_a \xi_b-\nabla_b \xi_a\right){h^c}_c + \left(\nabla_a {h^c}_b-\nabla_b{h^c}_a\right)\xi_c \\&+\left(\nabla_c \xi_a {h^c}_b - \nabla_c\xi_b {h^c}_a\right) - \left(\nabla_c {h^c}_b \xi_a - \nabla_c {h^c}_a \xi_b\right) \\
&- \left(\nabla_a {h^c}_c \xi_b - \nabla_b {h^c}_c \xi_a\right).
\end{split}
\end{equation}

We will take $\partial \Sigma$ for our case of a Schwarzschild-background to be a cross-section of the event horizon, so it is given by the coordinates $(v=\text{const.},r=\text{const.},x^A)$ and thus has topology of $S^2$ parameterized by the angular coordinates $x^A.$ In that case, \eqref{7} takes the form

\begin{equation}
\delta H_\xi \left[h_{ab}; g_{ab}\right] = -\frac{r^2}{16 \pi G} \oint_{\partial \Sigma} {d^2x \sqrt{\gamma} F_{rv}}.
\label{9}
\end{equation}

Evaluating \eqref{8} for our Schwarzschild-metric $g_{ab}$ in \eqref{4}, one obtains for \eqref{9}

\begin{equation} \label{10}
\begin{split}
&\delta H_\xi \left[h_{ab};g_{ab}\right] = -\frac{r^2}{16 \pi G} \oint_{\partial \Sigma} d^2x \sqrt{\gamma}\left( \right. \\
&\xi^v \left(-\frac{r_S}{2r^2}{h^A}_A-r^{-2}D^Ah_{Av}-\frac{2}{r}h_{vv}
-\frac{4}{r}\left(1-\frac{r_S}{r}\right)h_{vr}+\partial_v {h^A}_A \right. \\
&\left.- \left(1-\frac{r_S}{r}\right)r^{-2}D^Ah_{Ar}
-\frac{2}{r}\left(1-\frac{r_S}{r}\right)^2h_{rr}+\frac{1}{r}\left(1-\frac{r_S}{r}\right){h^A}_A \right. \\
&\left.+\left(1-\frac{r_S}{r}\right)\partial_r {h^A}_A\right) \\
&+\partial_r \xi^v \left(\frac{1}{2}\left(1-\frac{r_S}{r}\right)^2h_{rr}-\frac{1}{2}\left(1-\frac{r_S}{r}\right){h^A}_A+h_{vv}+\left(1-\frac{r_S}{r}\right)h_{vr} \right) \\
&+\partial_r \xi^r \left(-\frac{1}{2}\left(1-\frac{r_S}{r}\right)h_{rr}+\frac{1}{2}{h^A}_A\right) \\
&+\partial_v \xi^v \left(\frac{1}{2}\left(1-\frac{r_S}{r}\right)h_{rr}-\frac{1}{2}{h^A}_A\right) \\
&+\xi^r \left( \frac{r_S}{2r^2}h_{rr}+r^{-2}D^Ah_{Ar}+\frac{2}{r}h_{vr} +\frac{2}{r}\left(1-\frac{r_S}{r}\right)h_{rr} \right. \\
&\left.-\frac{1}{r}{h^A}_A-\partial_r{h^A}_A\right) \\
&+\xi^A \left(\partial_r h_{Av}-\frac{2}{r}h_{Av}-\partial_v h_{Ar}\right) \\
&\left.-\partial_v\xi^r h_{rr}+r^{-2}D^A\xi^v \left(h_{Av}+\left(1-\frac{r_S}{r}\right)h_{Ar}\right)-r^{-2}D^A\xi^r h_{Ar}\right).
\end{split}
\end{equation}

$D_A$ and $D^A$ denote the covariant derivative on the unit $2$-sphere where the index is raised and lowered with $\gamma_{AB}.$ 

Now we have derived the theoretical ground to accomplish our task. If we can find ``natural'' diffeomorphisms satisfying the two copies of Witt-algebra \eqref{3}, we are able to provide candidates for the Virasoro-generators of the black hole at the point $g_{ab} \in \Gamma$ in phase space.\footnote{Contrary to Carlip's approach, we do not impose the constraint of integrability on the diffeomorphisms $\xi_n$ and their generators \eqref{7}. This is because we allow the diffeomorphisms generated by the conformal generators to be field-dependent. Fortunately, the knowledge of those diffeomorphisms at the reference point $g_{ab} \in \Gamma$ given by \eqref{4} is sufficient to determine their generator algebra at this point. The purpose of the present analysis is to find out whether after all suited diffeomorphisms $\xi_n$ exist that could give rise to the conformal generators at the reference point $g_{ab} \in \Gamma.$ Away from $g_{ab} \in \Gamma,$ the diffeomorphisms generated by the conformal generators may look different and we will not determine them here. As a consequence, we are able to infer information about the conformal symmetry only right at the reference point $g_{ab}.$ For instance, we are not able to determine the temperature dependence in \eqref{42} away from $g_{ab}.$ We will come back to these issues in chapter \ref{Kapitel 5}.} Since the Hamiltonian generators $H_\xi$ have the dimension of an action, we can define dimensionless generators by 

\begin{equation} \label{11}
\begin{split}
H_{\xi_n} =: \hbar L_n \\
H_{\overline{\xi}_n} =: \hbar \overline{L}_n
\end{split}
\end{equation}

at $g_{ab} \in \Gamma$ and for $n \in \mathbb{Z}.$ Since the generators satisfy according to \eqref{1} and \eqref{3} two centrally extended Witt-algebras

\begin{equation} \label{12}
\begin{split}
\left\{H_{\xi_m}, H_{\xi_n} \right\} &= -i(m-n)H_{\xi_{m+n}} + K_{\xi_m,\xi_n} \\
\left\{H_{\overline{\xi}_m},H_{\overline{\xi}_n} \right\} &= -i(m-n)H_{\overline{\xi}_{m+n}} + K_{\overline{\xi}_m,\overline{\xi}_n} \\
\left\{ H_{\xi_m}, H_{\overline{\xi}_n} \right\} &= 0,
\end{split}
\end{equation}

we find after canonical quantization $\{ \cdot, \cdot\} \to \frac{1}{i \hbar}  [\cdot,\cdot ]$ of \eqref{12} that the Virasoro-generators $L_n$ and $\overline{L}_{n}$ fulfill two copies of the Virasoro-algebra in the standard form

\begin{equation} \label{13}
\begin{split}
\left[L_m,L_n\right] &= (m-n)L_{m+n}+\frac{c}{12}m(m^2-1)\delta_{m+n} \\
\left[\overline{L}_m,\overline{L}_n\right] &= (m-n)\overline{L}_{m+n}+\frac{\overline{c}}{12}m(m^2-1)\delta_{m+n} \\
\left[L_m, \overline{L}_n\right] &=0.
\end{split}
\end{equation}

$\delta_{m+n}=\delta_{m+n,0}$ denotes the Kronecker-delta. Thus, the central charges $c, \overline{c}$ and Virasoro-generators $L_0[g_{ab}], \overline L_0[g_{ab}]$ can be inferred from

\begin{equation} \label{14}
\begin{split}
\left. \delta_{\xi_{-m}}H_{\xi_m} \right|_{g_{ab}} &= -2i \hbar mL_0[g_{ab}]-i\frac{\hbar c}{12}m(m^2-1) \\
\left. \delta_{\overline \xi_{-m}} H_{\overline \xi_m} \right|_{g_{ab}} &= -2i \hbar m\overline L_0[g_{ab}]-i\frac{\hbar \overline c}{12}m(m^2-1)
\end{split}
\end{equation}

for $m \in \mathbb{Z}.$ 

Equation \eqref{14} already fixes the data needed for the counting of state degeneracy in a CFT. The computation of the left hand side of \eqref{14} can be done by \eqref{10}.

To summarize, the task left for the next chapter is to find a ``preferred'' Witt-algebra of diffeomorphisms \eqref{3}. We will have to argue in what sense these diffeomorphisms will be preferred. But if this can be accomplished, the associated Hamiltonian generators will provide natural Virasoro-generators of a possible CFT describing part of the phase space responsible for the microstates of the black hole.

Nevertheless, the choice of diffeomorphisms is at this stage only a guess. It might be that chosen diffeomorphisms have nothing to do with the symmetry generators of the aforementioned CFT. Within this approach, it is even not clear that such a CFT exists. However, in this chapter we have shown that the choice of diffeomorphisms \eqref{3} fixes via equations \eqref{14} and \eqref{10} the central charges and conformal weights (or equivalently the temperatures) of a would-be CFT. This is already enough data to determine the degeneracy of states in this CFT in order to see whether it agrees with the Bekenstein-Hawking entropy of the black hole. 

\section{Searching for Virasoro-Algebra}
\label{Kapitel 3}

The goal of this chapter is to find out, whether the presence of a black hole event horizon singles out a Witt-algebra of diffeomorphisms \eqref{3} in a natural way and what natural in this context might mean. In the last chapter, we have explained how such diffeomorphisms could be related to the generators of a conformal symmetry governing the black hole's phase space and provided formulas to extract information of this CFT directly from the diffeomorphisms. 

\subsection{Virasoro-Vectorfields}
\label{Kapitel 3.1}

Given the Schwarzschild black hole $g_{ab} \in \Gamma$ in \eqref{4}, what diffeomorphisms forming a Witt-algebra might be singled out? Remember, that a Witt-algebra \eqref{2} is isomorphic to the algebra $\mathfrak{diff \ } S^1$ of all diffeomorphisms on $S^1.$ The question can thus be rephrased as whether there are preferred directions in a Schwarzschild-spacetime. If so, periodic reparameterizations along these directions provide a Witt-algebra $\mathfrak{diff \ } S^1$ and the periodicities would then fix the temperatures of a possible CFT. In \cite{Carlip:1999cy} Carlip provided for the general case of a local Killing horizon a candidate for such a preferred direction. Although the associated Hamiltonian generators can be shown to generate central extensions and to give rise to the correct entropy, only one copy of a Virasoro-algebra is found. 

Instead, we want to give a somewhat different proposal for constructing diffeomorphisms forming the algebra \eqref{3}. We will see that the associated Hamiltonian generators will indeed form two copies of Virasoro-algebra with equal central charges as one would expect for the symmetry algebra of a CFT. The diffeomorphisms we are going to construct were already given in \cite{Averin:2018owq} up to cosmetic changes. We now explain how they are singled out. 

We keep the philosophy of \cite{Carlip:1999cy} that a local Killing horizon singles out a preferred direction which provides the basis for the construction of a $\mathfrak{diff \ } S^1$ algebra. However, a Schwarzschild geometry has in its maximal extension two event horizons and we propose to treat both on the same footing in the search for an algebra \eqref{3}. 

In the well-known Kruskal-coordinates $(U,V,x^A)$ which cover the entire maximal extension of the Schwarzschild spacetime, the metric takes the form

\begin{equation}
ds^2 = -\frac{4r_S^3}{r}e^{-\frac{r}{r_S}}dUdV + r^2\gamma_{AB}dx^Adx^B.
\label{15}
\end{equation}

The future (past) event horizon is located at $U=0$ $(V=0).$ Indeed, the Schwarzschild geometry \eqref{15} provides the \emph{two} preferred lightlike directions $\partial_U$ and $\partial_V.$ 

However, the Cardy-formula provides the degeneracy of states $S=S(L_0,\overline L_0)$ in a CFT at particular values of the Virasoro-generators $L_0$ and $\overline L_0.$ Since the black hole entropy gives the state degeneracy at fixed mass and angular momentum, the candidates for the Virasoro zero-modes $H_{\xi_0}$ and $H_{\overline \xi_0}$ should therefore ``measure'' the mass and angular momentum parameter of the black hole. Therefore, it seems that reparameterizations along the direction $U$ $(V)$ are not enough. The diffeomorphisms $\xi_n$ and $\overline \xi_n$ to form \eqref{3} should also contain components in the direction $\partial_\phi$ which is conjugated to the angular momentum.\footnote{One can also check that although reparameterizations along $U$ $(V)$ form the algebra \eqref{3}, the Hamiltonian generators do not develop a central charge in \eqref{14}.}

Thus, in Kruskal-coordinates $(U,V,\theta,\phi),$ we make the ansatz for the vectorfields \eqref{3}

\begin{equation} \label{16}
\begin{split}
\xi_n &= f_n \partial_V + g_n \partial_\phi \\
\overline \xi_n &= \overline f_n \partial_U + \overline g_n \partial_\phi
\end{split}
\end{equation}

for $n \in \mathbb{Z}.$ Here, the functions

\begin{equation} \label{17}
\begin{split}
f_n &= f_n(V,\phi) \\
g_n &= g_n(V,\phi) \\
\overline f_n &= \overline f_n(U,\phi) \\
\overline g_n &= \overline g_n(U,\phi)
\end{split}
\end{equation}

need to be determined from the requirement \eqref{3}. The first equation in \eqref{3} yields two conditions on the functions \eqref{17}, namely

\begin{equation}
f_m\partial_Vf_n + g_m\partial_\phi f_n - \left(m \longleftrightarrow n\right) = -i(m-n)f_{m+n}
\label{18}
\end{equation}

and

\begin{equation}
f_m\partial_Vg_n+g_m\partial_\phi g_n-\left(m \longleftrightarrow n\right) = -i(m-n)g_{m+n}
\label{19}
\end{equation}

for $m,n \in \mathbb{Z}.$ Analogous equations follow from the second equation in \eqref{3} for the anti-chiral functions $\overline f_n, \overline g_n.$ The last equation of \eqref{3} then yields the conditions

\begin{equation} \label{20}
\begin{split}
\partial_\phi \overline f_n &= 0\\
\partial_\phi f_n &= 0\\
g_m\partial_\phi\overline g_n - \overline g_n \partial_\phi g_m &= 0.
\end{split}
\end{equation}

The last conditions of \eqref{20} can be fulfilled by choosing the product ansatz

\begin{equation} \label{21}
\begin{split}
g_m(V,\phi) &= \Phi(\phi) G_m(V)\\
\overline g_m(U,\phi) &= \Phi(\phi) \overline G_m(U).
\end{split}
\end{equation}

With these restrictions on $f_n$ and $g_n,$ equation \eqref{18} becomes 

\begin{equation}
f_m\partial_V f_n-\left(m \longleftrightarrow n\right) = -i(m-n)f_{m+n}
\label{22}
\end{equation}

and \eqref{19} yields

\begin{equation}
f_m\partial_V G_n-\left(m \longleftrightarrow n\right) = -i(m-n)G_{m+n}.
\label{23}
\end{equation}

Choosing $f_n=f_n(V)$ to satisfy \eqref{22}, equation \eqref{23} is fulfilled with the choice

\begin{equation}
G_n = \partial_V f_n.
\label{24}
\end{equation}

Therefore, the vectorfields 

\begin{equation} \label{25}
\begin{split}
\xi_n &= f_n(V)\partial_V + \Phi(\phi)\partial_V f_n\partial_\phi \\
\overline \xi_n &= \overline f_n(U)\partial_U + \Phi(\phi) \partial_U \overline f_n \partial_\phi
\end{split}
\end{equation}

provide an algebra \eqref{3} if the functions $f_n=f_n(V)$ are chosen to satisfy \eqref{22} and $\overline f_n$ are chosen analogously. $\Phi=\Phi(\phi)$ is at this stage arbitrary. A legal choice is then

\begin{equation} \label{26}
\begin{split}
f_n(V) &= \frac{1}{A} V^{1+inA} \\
\overline f_n(U) &= \frac{1}{B} U^{1+inB}
\end{split}
\end{equation}

where $A, B \in \mathbb{R} \backslash \{0\}$ are so far arbitrary parameters. 

Unfortunately, the constructed vectorfields \eqref{25} are still not satisfactory. In order to give rise to independent Virasoro generators $L_n,$ the $\xi_n$ have to be linearly independent functions of the angular coordinates. One possibility is that a factor $e^{in\phi}$ appears in \eqref{25} instead of a fixed function $\Phi(\phi).$ However, this is now easy to achieve. Since \eqref{25} satisfies a Witt-algebra \eqref{3}, we can generate such vectorfields out of \eqref{25} by applying an active coordinate transformation. The new vectorfields then still satisfy \eqref{3}. We choose the coordinate transformation\footnote{The vectorfields of \cite{Averin:2018owq} are obtained by putting an additional minus sign in the exponential of the first equation in \eqref{27}. This was omitted here in order to make the frequencies and temperatures positive that are going to appear later. In addition, we replaced $A,B$ with their inverse values as compared to \cite{Averin:2018owq}.}

\begin{equation} \label{27}
\begin{split}
U' &= Ue^{\frac{1}{B}\phi} \\
V' &= Ve^{-\frac{1}{A}\phi} \\
\theta &= \theta \\
\phi &= \phi.
\end{split}
\end{equation}

The procedure yields

\begin{equation}
\xi_n^a=
\begin{pmatrix}
\xi_n^U\\
\xi_n^V\\
\xi_n^\theta\\
\xi_n^\phi
\end{pmatrix}
=
\begin{pmatrix}
\frac{1}{AB}U(1+inA)V^{inA}e^{in\phi}\Phi(\phi)\\
\frac{1}{A}V^{1+inA}e^{in\phi}-\frac{1}{A^2}(1+inA)V^{1+inA}e^{in\phi}\Phi(\phi)\\
0\\
\frac{1}{A}(1+inA)V^{inA}e^{in\phi}\Phi(\phi)
\end{pmatrix}
\label{28}
\end{equation}

and

\begin{equation}
\overline \xi_n^a=
\begin{pmatrix}
\frac{1}{B}U^{1+inB}e^{-in\phi}+\frac{1}{B^2}(1+inB)U^{1+inB}e^{-in\phi}\Phi(\phi)\\
-\frac{1}{AB}V(1+inB)U^{inB}e^{-in\phi}\Phi(\phi)\\
0\\
\frac{1}{B}(1+inB)U^{inB}e^{-in\phi}\Phi(\phi)
\end{pmatrix}
.
\label{29}
\end{equation}
 
In order to meet the conventions of the last chapter, we formulate \eqref{28} and \eqref{29} in Eddington-Finkelstein coordinates $(v,r,\theta,\phi)$ getting

\begin{equation} \label{30}
\begin{split}
\xi_n^a=
\begin{pmatrix}
\xi_n^v\\
\xi_n^r\\
\xi_n^\theta\\
\xi_n^\phi
\end{pmatrix}
=
\begin{pmatrix}
2r_S\left(\frac{1}{A}-\frac{1}{A^2}(1+inA)\Phi(\phi)\right)\\
\left(1-\frac{r_S}{r}\right)r_S\left(\frac{1}{A}+\frac{1}{A}\left(-\frac{1}{A}+\frac{1}{B}\right)(1+inA)\Phi(\phi)\right)\\
0\\
\frac{1}{A}(1+inA)\Phi(\phi)
\end{pmatrix}
\times \\
e^{inA\frac{v}{2r_S}}e^{in\phi}
\end{split}
\end{equation}

and

\begin{equation}
\begin{split}
\overline \xi_n^a=
\begin{pmatrix}
-2r_S\frac{1}{AB}(1+inB)\Phi(\phi)\\
r_S\left(1-\frac{r_S}{r}\right)\left(\frac{1}{B}+\frac{1}{B}\left(-\frac{1}{A}+\frac{1}{B}\right)(1+inB)\Phi(\phi)\right)\\
0\\
\frac{1}{B}(1+inB)\Phi(\phi)
\end{pmatrix}
\times \\
(-1)^{inB}e^{inB\frac{r^*}{r_S}}e^{-inB\frac{v}{2r_S}}e^{-in\phi}
\end{split}
\label{31}
\end{equation}

for $n \in \mathbb{Z}$ and with $r^*$ from \eqref{6}. These vectorfields fulfill $\left(\xi_n^a\right)^*=\xi_{-n}^a$ and $\left( \overline \xi_n^a \right)^*=\overline \xi_{-n}^a$ and form two copies of Witt-algebra \eqref{3} as required. Formulas \eqref{30} and \eqref{31} are the main result of this chapter and will be used in the next chapter for the entropy counting within the framework developed in chapter \ref{Kapitel 2}. In what follows, we will try to fix the remaining arbitrary function $\Phi(\phi)$ and $A,B \in \mathbb{R} \backslash \{0\}.$ On the road, we will also comment on different approaches made to find such Virasoro-vectorfields. 

\subsection{Temperatures}
\label{Kapitel 3.2}

Although our construction of \eqref{30} and \eqref{31} is motivated by \cite{Carlip:1999cy}, we note that our final result is really different. Our vectorfields violate the horizon boundary conditions proposed in \cite{Carlip:1999cy} and thus the construction is genuinely different. This is mainly due to the appearance of a $\partial_\phi$-component in our choice of diffeomorphisms. Indeed, we wanted this component to appear in order for the Virasoro zero-modes to ``measure'' the black hole's mass and angular momentum parameter. The Virasoro zero-modes are according to chapter \ref{Kapitel 2} induced by the $\mathfrak{u}(1) \oplus \overline{\mathfrak{u}(1)}$-subalgebra of \eqref{3} spanned by $\xi_0$ and $\overline \xi_0.$ These vectorfields hence should be - in Schwarzschild-coordinates $(t,r,\theta,\phi)$ - linear combinations of $\partial_t$ and $\partial_\phi.$ For this to be fulfilled, in \eqref{30} and \eqref{31} the $r$-component has to vanish for $n=0$ which requires the choice

\begin{equation}
\Phi(\phi) = \frac{AB}{B-A}.
\label{32}
\end{equation}

With this choice the $\mathfrak{u}(1) \oplus \overline{\mathfrak{u}(1)}$-subalgebra is given by 

\begin{equation} \label{33}
\begin{split}
\xi_0 &= -2r_S\frac{1}{B-A}\partial_t+\frac{B}{B-A}\partial_\phi \\
\overline \xi_0 &= -2r_S\frac{1}{B-A}\partial_t+\frac{A}{B-A}\partial_\phi.
\end{split}
\end{equation}

By some thermodynamic considerations, \eqref{33} provides enough information to fix the temperatures of the CFT described in chapter \ref{Kapitel 2} that would be associated with the full Witt-algebra \eqref{3}. However, note that these considerations provide rather a consistency check as the temperatures are already fixed through \eqref{14} by $L_0[g_{ab}]$ and $\overline L_0[g_{ab}]$: The temperatures inferred from the algebra \eqref{14} can be compared to the temperatures that are thermodynamically obtained from \eqref{33} by a procedure to be explained in the following. 

In \cite{Haco:2018ske}, a different approach to find Virasoro-vectorfields was made but the latter consistency check was not done. There, the $im$-terms which would determine the temperature are neither displayed in equation $(5.15)$ nor in the counterterm-correction $(5.16)$. It would be interesting to see whether in \cite{Haco:2018ske} the temperatures obtained algebraically via \eqref{14} agree with the temperatures inferred thermodynamically. 

In order to thermodynamically infer the CFT temperatures, we introduce a scalar field to be put in thermal contact with the black hole. Consider a free massless Klein-Gordon field on the Schwarzschild-background. Its eigenmodes are of the form $F(r,\theta)e^{-i\omega t+im\phi}$ with frequeny $\omega$ and angular momentum $m.$ These are then also eigenfunctions of \eqref{33} with eigenvalues

\begin{equation} \label{34}
\begin{split}
\xi_0 &= -in = -i\left(\frac{2r_S\omega}{A-B}-m\frac{B}{B-A}\right) \\
\overline \xi_0 &= -i\overline n = -i\left(\frac{2r_S\omega}{A-B}-m\frac{A}{B-A}\right).
\end{split}
\end{equation}

If we would allow for backreaction, the scalar field can exchange energy and angular momentum with the gravitational field under the constraint that both are conserved in total. In thermal equilibrium, for the Schwarzschild black hole, the scalar eigenmodes $(\omega,m)$ are thermally distributed weighted by a Boltzmann-factor

\begin{equation*}
e^{-\frac{\hbar \omega}{T_H}}
\end{equation*}

with the Hawking-temperature $T_H=\frac{\hbar}{4 \pi r_S}.$ This is rewritten in terms of the eigenfrequencies $(n, \overline n)$ as $e^{-\frac{n}{T}-\frac{\overline n}{\overline T}}$ with the temperatures

\begin{equation} \label{35}
\begin{split}
T &= \frac{1}{2 \pi} \frac{1}{A} \\
\overline T &= -\frac{1}{2 \pi} \frac{1}{B}.
\end{split}
\end{equation}

Due to the zeroth law of thermodynamics, \eqref{35} are also the temperatures of the CFT governing the black hole if associated to the diffeomorphisms \eqref{30} \eqref{31}. 

\subsection{$SL(2;\mathbb{R})$-Casimir and Conformal Symmetry in Scattering}
\label{Kapitel 3.3}

In this chapter, we have presented one particular way to single out Virasoro-vectorfields \eqref{30},\eqref{31} which we will use for entropy counting via Carlip's approach in the next chapter. The recent work \cite{Haco:2018ske} also follows Carlip's approach but a different philosophy is used to find a conformal algebra of vectorfields.\footnote{The analysis there is done for a Kerr black hole (Kerr-Newman in \cite{Haco:2019ggi}) and diverges in the Schwarzschild-limit but this will be not important here.}

The Witt-algebra \eqref{3} has a $\mathfrak{sl}(2,\mathbb{R}) \oplus \overline{\mathfrak{sl}(2,\mathbb{R})}$-subalgebra spanned by $\xi_{-1},\xi_0,\xi_1$ and its anti-chiral counterpart. Associated to this global conformal algebra are the Casimir-operators

\begin{equation}
\mathcal{H}^2 = -\mathcal L_{\xi_0} \mathcal L_{\xi_0} + \frac{1}{2}\left(\mathcal L_{\xi_1} \mathcal L_{\xi_{-1}}+\mathcal L_{\xi_{-1}} \mathcal L_{\xi_1}\right)
\label{36}
\end{equation}

and an analogous anti-chiral expression. Now, one can try to find a preferred $\mathfrak{sl}(2,\mathbb{R})$-algebra of diffeomorphisms by studying the form of the associated differential operator \eqref{36}. In \cite{Castro:2010fd} vectorfields forming an $\mathfrak{sl}(2,\mathbb{R}) \oplus \overline{\mathfrak{sl}(2,\mathbb{R})}$-algebra were given. It was further shown there, that for a free massless Klein-Gordon field in a Kerr-background - in a suited regime - eigenfunctions of \eqref{36} give rise to eigenmodes of the Klein-Gordon equation. As a consequence of the $\mathfrak{sl}(2,\mathbb{R}) \oplus \overline{\mathfrak{sl}(2,\mathbb{R})}$-invariance of \eqref{36}, scattering - in a suited in regime - behaves as being invariant under a ``hidden'' 2D global conformal symmetry (see \cite{Castro:2010fd} for further details).

The idea of \cite{Haco:2018ske} is then to find a full local conformal $Vir \oplus \overline{Vir}$-algebra \eqref{3} of diffeomorphisms which realizes the latter hidden conformal symmetry and then can be used for entropy counting.  

However, to our understanding the $Vir \oplus \overline{Vir}$-vectorfields proposed in \cite{Haco:2018ske} form only an enhancement of the $\mathfrak u(1) \oplus \overline{\mathfrak u(1)}$-algebra given in \cite{Castro:2010fd} (spanned by $\xi_0,\overline \xi_0$). They seem not to contain the global conformal $\mathfrak{sl}(2,\mathbb{R}) \oplus \overline{\mathfrak{sl}(2,\mathbb{R})}$-algebra of \cite{Castro:2010fd} (spanned by $\xi_n, \overline \xi_n$ for $n=-1,0,1$).

But then, we do not understand in what way the Kerr-geometry singles out the $Vir \oplus \overline{Vir}$-vectorfields of \cite{Haco:2018ske} so that their Hamiltonian generators could govern a possible CFT of the Kerr black hole. Indeed, in \cite{Haco:2018ske}, the expected central charges are only obtained from the generator algebra after non-canonical counterterm-corrections. 

Nevertheless, it might still be useful to have the $\mathfrak{sl}(2,\mathbb{R})$-Casimir \eqref{36} in mind. What is its meaning for our $Vir \oplus \overline{Vir}$ choice in \eqref{30},\eqref{31}? Inspired by \cite{Castro:2010fd} and our construction of these diffeomorphisms, a natural expectation would be that $\mathcal H^2$ and $\overline{\mathcal H}^2$ possibly govern the scalar scattering in the vicinity of the bifurcation of the horizons in a suited regime of parameters. We do not enter an analysis of these questions further at this place. However, we want to note that a computation reveals that the expression for \eqref{36} with the vectorfields \eqref{30}\eqref{31} indeed greatly simplyfies for the choice

\begin{equation}
B=-A.
\label{37}
\end{equation}

This could be a hint that the $Vir$-algebra \eqref{30} and $\overline{Vir}$-algebra \eqref{31} can belong to the Virasoro-algebra of the same CFT only with the choice \eqref{37}. However, we will leave the parameters $A$ and $B$ unspecified. We will then see further evidence for this conjecture from the fact, that the Cardy-entropy will get extremized precisely for the choice \eqref{37}. 

Using \eqref{32} and \eqref{37}, the last unspecified parameter in \eqref{30} and \eqref{31} is then $A.$ We will see that it will cancel out of the entropy counting. Such an ambiguity parameter was already present in \cite{Carlip:1998wz}. As explained there, euclidean quantum gravity suggests the choice $A=1$ together with \eqref{37} since the wavenumber for $v$ is then given by the surface gravity $\kappa=\frac{1}{2r_S}.$ However, we will leave $A$ unspecified since this ambiguity can have a mathematical meaning as can be seen in chapter \ref{Kapitel 5}. 

\section{Entropy Counting}
\label{Kapitel 4}

With the vectorfields \eqref{30} and \eqref{31} of the last chapter we are now ready to apply the framework of chapter \ref{Kapitel 2} for entropy counting. 

\subsection{Schwarzschild-Entropy}
\label{Kapitel 4.1}

Our goal is to determine the data on the right hand side of \eqref{14}. To this end, we evaluate the left hand side of \eqref{14} using the vectorfields determined in \eqref{30} and \eqref{31}. A computation yields

\begin{equation} \label{38}
\begin{split}
& \left. \delta_{\xi_{-m}}H_{\xi_m}\right|_{g_{ab}} \\
=&-im\frac{r^2}{4G}\left(\frac{r_S^2}{r^2}\frac{1}{A}\Phi\left(\frac{2}{A}-\frac{2}{A}\left(\frac{1}{A}+\frac{1}{B}\right)\Phi\right) \right. \\
& \left. +\frac{r_S}{r}\left(1-\frac{r_S}{r}\right)\left(-\frac{4}{A}-\frac{4}{AB}\Phi+\frac{4}{A^2}\Phi+\frac{8}{A^2B}\Phi^2\right)\right) \\
&-im^3\frac{r^2}{4G}\left(2\Phi-2\frac{r_S^2}{r^2}\Phi^2\left(\frac{1}{A}+\frac{1}{B}\right)+\frac{r_S}{r}\left(1-\frac{r_S}{r}\right)\frac{8}{B}\Phi^2\right) \\
&+im^3\frac{r_S^2}{8G}\int_0^\pi {d\theta \sin^{-1}(\theta) \left(1-\frac{r_S}{r}\right)\frac{8}{AB}\Phi}.
\end{split}
\end{equation}

In the limit $r \to r_S$ the expression is well-defined and we get from \eqref{14}

\begin{equation} \label{39}
\begin{split}
L_0[g_{ab}]-\frac{c}{24} &= \frac{r_S^2}{4 \hbar G}\left(\frac{1}{A}\Phi\left(\frac{1}{A}-\frac{1}{A}\left(\frac{1}{A}+\frac{1}{B}\right)\Phi \right)\right) \\
c &= \frac{3r_S^2}{\hbar G}\left(2\Phi-2\Phi^2\left(\frac{1}{A}+\frac{1}{B}\right)\right).
\end{split}
\end{equation}

Using \eqref{32} and applying the Cardy-formula\footnote{We take the convention $A>0$ and $B<0$ in the following in order for the temperatures \eqref{35} to be positive.}

\begin{equation}
S_{chiral}=2\pi\sqrt{\frac{c}{6}\left(L_0-\frac{c}{24}\right)}=\frac{\pi r_S^2}{\hbar G}\frac{-2AB}{(B-A)^2}.
\label{40}
\end{equation}

Consistently, the canonical version of the Cardy-formula with the temperatures \eqref{35} yields the same result

\begin{equation}
S_{chiral}=\frac{\pi^2}{3}cT=\frac{\pi r_S^2}{\hbar G}\frac{-2AB}{(B-A)^2}.
\label{41}
\end{equation}

As conjectured in chapter \ref{Kapitel 3.3}, these expressions are maximized if \eqref{37} holds. In this case, one has

\begin{equation} \label{42}
\begin{split}
L_0[g_{ab}]-\frac{c}{24} &= \frac{r_S^2}{8 \hbar G}\frac{1}{A} \\
c &= \frac{3r_S^2}{\hbar G}A
\end{split}
\end{equation}

and

\begin{equation}
S_{chiral}=\frac{1}{2}\frac{\pi r_S^2}{\hbar G}.
\label{43}
\end{equation}

The anti-chiral contribution is determined in the same way 

\begin{equation} \label{44}
\begin{split}
&\left. \delta_{\overline \xi_{-m}}H_{\overline \xi_m}\right|_{g_{ab}} \\
=&-im\frac{r^2}{4G}\left(\frac{r_S^2}{r^2}\left(\frac{2}{B^2}\Phi+2\left(\frac{1}{A}+\frac{1}{B}\right)\frac{1}{B^2}\Phi^2\right) \right. \\
& \left. -\frac{r_S}{r}\left(1-\frac{r_S}{r}\right)\left(2\Phi\left(\frac{2}{AB}-\frac{2}{B^2}\right)-\frac{4}{B}+\Phi^2\frac{8}{AB^2}\right)\right) \\
&-im^3 \frac{r^2}{4G}\left(2\Phi+2\frac{r_S^2}{r^2}\left(\frac{1}{A}+\frac{1}{B}\right)\Phi^2-\frac{r_S}{r}\left(1-\frac{r_S}{r}\right)\Phi^2\frac{8}{A}\right) \\
&+im^3\frac{r_S^2}{8G}\int_0^\pi {d\theta \sin^{-1}(\theta)\left(1-\frac{r_S}{r}\right)\frac{8\Phi}{AB}}.
\end{split}
\end{equation}

In the limit $r \to r_S$ the expression is well-defined and we get from \eqref{14}

\begin{equation} \label{45}
\begin{split}
\overline L_0[g_{ab}]-\frac{\overline c}{24} &= \frac{r_S^2}{4 \hbar G}\left(\frac{1}{B^2}\Phi+\left(\frac{1}{A}+\frac{1}{B}\right)\frac{1}{B^2}\Phi^2\right) \\
\overline c &= \frac{3r_S^2}{\hbar G}\left(2\Phi+2\left(\frac{1}{A}+\frac{1}{B}\right)\Phi^2\right).
\end{split}
\end{equation}

Using \eqref{32} and applying the Cardy-formula 

\begin{equation}
S_{anti-chiral}=2\pi\sqrt{\frac{\overline c}{6}\left(\overline L_0-\frac{\overline c}{24}\right)}=\frac{\pi r_S^2}{\hbar G}\frac{-2AB}{(B-A)^2}.
\label{46}
\end{equation}

Consistently, the canonical version of the Cardy-formula with the temperatures \eqref{35} yields the same result

\begin{equation}
S_{anti-chiral}=\frac{\pi^2}{3}\overline c \overline T=\frac{\pi r_S^2}{\hbar G}\frac{-2AB}{(B-A)^2}.
\label{47}
\end{equation}

These expressions coincide with the chiral contribution and are also maximized if \eqref{37} is fulfilled. For that case, one has

\begin{equation} \label{48}
\begin{split}
\overline L_0[g_{ab}]-\frac{\overline c}{24} &= \frac{r_S^2}{8 \hbar G}\frac{1}{A} \\
\overline c &= \frac{3r_S^2}{\hbar G}A
\end{split}
\end{equation}

and

\begin{equation}
S_{anti-chiral}=\frac{1}{2}\frac{\pi r_S^2}{\hbar G}.
\label{49}
\end{equation}

Thus, the total Cardy-entropy is

\begin{equation}
S=S_{chiral}+S_{anti-chiral}=\frac{\pi r_S^2}{\hbar G}.
\label{50}
\end{equation}

This matches precisely the Bekenstein-Hawking entropy of a Schwarzschild black hole.

\subsection{Extrapolation to the General Case}
\label{Kapitel 4.2}

The CFT data derived in \eqref{42}, \eqref{48} and \eqref{35} can be written in the form

\begin{equation} \label{51}
\begin{split}
c&=\overline c=\frac{3 \mathcal{A}}{4\pi \hbar G}A \\
T&=\overline T=\frac{1}{2 \pi}\frac{1}{A}
\end{split}
\end{equation}

with the horizon area $\mathcal{A}.$ In \cite{Carlip:2011ax} one copy of a Witt-algebra of vectorfields was presented to reproduce central charges and temperatures similar to the chiral half of \eqref{51} for the general case of a stationary black hole of dimension $3+1.$ However, there are some differences. In \cite{Carlip:2011ax} these quantities contain divergences which cancel out in entropy counting and the temperature is derived only by thermodynamic considerations and not from a computation of Virasoro zero-modes. In addition, the chiral Virasoro-algebra in \cite{Carlip:2011ax} is only able to account for half of the expected Bekenstein-Hawking entropy. 

For a Schwarzschild black hole, we have managed to provide the missing second copy of Virasoro-algebra accounting for the second missing half of the entropy. In addition, our choice of $Vir \oplus \overline{Vir}$-vectorfields leads to well-defined quantities \eqref{51} that contain no divergences. Furthermore, the temperatures in \eqref{51} are consistently in agreement with their derivation from Virasoro zero-modes $L_0[g_{ab}]$ and $\overline L_0[g_{ab}]$ using covariant phase space methods.

We derived \eqref{51} for the case of a Schwarzschild black hole. However, our methods employed allow for canonical generalization. The strategy to pick out a $Vir \oplus \overline{Vir}$-algebra of vectorfields can be analogously applied in the general case. Due to the similarities of \eqref{51} and the general analysis of \cite{Carlip:2011ax}, we conjecture \eqref{51} to apply also in this general case leading to the entropy

\begin{equation}
S_{Cardy}=\frac{\pi^2}{3}cT+\frac{\pi^2}{3}\overline c \overline T=\frac{\mathcal{A}}{4 \hbar G}
\label{52}
\end{equation}

as required. Note, that we have provided a proof of \eqref{51} and \eqref{52} only for the Schwarzschild case and left the general case as a conjecture. Checking the conjecture would now require a straightforward computation that we do not enter at this place. 

\section{Discussion and Interpretation}
\label{Kapitel 5}

In the preceding chapters, we have revisited Carlip's approach to entropy counting. We have provided a $Vir \oplus \overline{Vir}$-algebra of diffeomorphisms and analyzed the algebra of the associated Hamiltonian generators. We found that the latter give rise to a Virasoro-algebra such that counting the state degeneracy of the would-be CFT is in agreement with Bekenstein-Hawking entropy. 

So far, this approach does not tell much about this would-be CFT that possibly governs the part of phase space responsible for black hole microstates. What is needed, is to analyze the Hamiltonian phase space in the vicinity of a black hole state in a systematic fashion. In \cite{Averin:2018owq} a systematic way was proposed to analyze the Hamiltonian phase space of general relativity and to find a dual theory describing the relevant part of phase space responsible for black hole microstates. 

Here, we want to briefly sketch how and why such a systematic treatment works in order to show how the entropy counting presented here fits into this procedure. An application of this treatment including the role of Carlip's entropy counting was already given in \cite{Averin:2018owq}. A more detailed description of the treatment itself including applications to simpler theories than gravity will be provided somewhere else. Instead, here we will just sketch the main ideas. 

\subsection{Holography in Covariant Phase Space}
\label{Kapitel 5.1}

Consider an arbitrary field theory over some $n$-dimensional manifold $M$ given by an action $S=S[\Phi].$ We denote the fields in the theory collectively by $\Phi.$ The goal is to analyze the Hamiltonian phase space in a structured way. Due to its flexibility, we use the covariant phase space approach for our explanations \cite{Lee:1990nz,Wald:1999wa} (see \cite{Seraj:2016cym} for a review). The main idea of the covariant phase space approach is the observation that the Hamiltonian phase space is isomorphic to the set of all field configurations satisfying the field equations. This solution space $\overline{\mathcal{F}}$ is equipped with a suited (pre)symplectic form and (after dividing out symplectic zero-modes through suitable gauge-fixing) then gives rise to the covariant phase space $\Gamma$ equivalent to the Hamiltonian phase space. 

We denote the coordinates on the covariant phase space by $[\Phi]^A$ with $A,B,\ldots$ being the indexes. The action $S$ is assigned a differential form $\omega=\omega[\delta_1 \Phi,\delta_2 \Phi;\Phi]$ of degree $n-1$ over $M$ which is in addition a closed $2$-form over the space of all field configurations. On shell, that is for $\Phi \in \overline{\mathcal{F}}$ and $\delta_1 \Phi,\delta_2 \Phi \in T_{\Phi} \overline{\mathcal{F}},$ $\omega$ is exact $\omega=dk$ for a form $k=k[\delta_1 \Phi,\delta_2 \Phi;\Phi]$ of degree $n-2$ over $M.$ 

Let $\Sigma \subseteq M$ be a hypersurface with the boundary $\partial \Sigma=B_1 \cup B_2,$ where $B_1$ and $B_2$ are disconnected codimension $2$ surfaces. The symplectic flow passing through $\Sigma$ is then on-shell given by a boundary integral

\begin{equation}
\int_\Sigma {\omega[\delta_1 \Phi,\delta_2 \Phi;\Phi]} = \oint_{B_1} {k[\delta_1 \Phi,\delta_2 \Phi;\Phi]}-\oint_{B_2} {k[\delta_1 \Phi,\delta_2 \Phi;\Phi]}.
\label{53}
\end{equation}

Each of the boundary integrals in \eqref{53} can be used to define a symplectic form over $\Gamma.$ 

\begin{equation}
\oint_{B_1} {k[\delta_1 \Phi, \delta_2 \Phi;\Phi]} = \Omega_{AB} \left[\delta_1 \Phi\right]^A \left[ \delta_2 \Phi\right]^B
\label{54}
\end{equation}

defines the symplectic form $\Omega_{AB}=\Omega_{AB}^{(B_1)}$ over $\Gamma$ (relative to $B_1$). This can then be used to define the Poisson-bracket in the usual way. We denote quantities sometimes with the superscript $(B_1)$ to remember that they are defined with $B_1$ as the chosen reference. For a vectorfield $X$ over $\Gamma$ corresponding to field variations $\delta_X \Phi,$ the expression 

\begin{equation}
\delta H_X\left[\delta \Phi;\Phi\right]=\oint_{B_1} {k\left[\delta \Phi,\delta_X \Phi;\Phi\right]}
\label{55}
\end{equation}

defines a $1$-form over the phase space. If $X$ is a symplectic symmetry $\mathcal L_X \Omega_{AB}^{(B_1)}=0,$ the $1$-form \eqref{55} is exact and can be integrated over phase space to provide the scalar $H_X=H_X^{(B_1)}[\Phi].$ The role of this scalar is to generate $\delta_X \Phi$ via the Poisson-bracket. 

Due to the expression \eqref{55} the value $H_X[\Phi]$ contains information about the field configuration $\Phi$ over the surface $B_1.$ In fact, for the linearly independent symplectic symmetries $X,$ we can think of $H_X[\Phi]$ as part of the Cauchy-data required to specify $\Phi \in \Gamma.$ These values $H_X[\Phi]$ can therefore be thought of forming part of the coordinates of a chart for the phase space $\Gamma.$ Due to their holographic nature \eqref{55}, we termed them in \cite{Averin:2018owq} as boundary Cauchy-data (BCD). 

The BCD is defined with respect to the codimension $2$ surface $B_1.$ What would have been if we had wanted to define it with repect to a different surface $B_2$ of codimension $2?$ In that case, we have to choose a hypersurface $\Sigma$ connecting $B_1$ and $B_2$ and correct the BCD relative to $B_1$ by the symplectic current passing through $\Sigma.$ Due to \eqref{53} and \eqref{55} the BCD of a symplectic symmetry $X$ are related by

\begin{equation}
\delta H_X^{(B_1)}\left[\delta \Phi;\Phi\right]-\delta H_X^{(B_2)}\left[\delta \Phi;\Phi\right]=\int_\Sigma {\omega\left[\delta \Phi,\delta_X \Phi;\Phi\right]}.
\label{56}
\end{equation}

That is, the change of the BCD from a surface $B_1$ to a surface $B_2$ is dictated by the symplectic current $\omega[\delta \Phi,\delta_X \Phi;\Phi]$ passing through the hypersurface in between. The specification of these symplectic currents along an entire Cauchy-surface $\Sigma$ forms the remaining Cauchy-data (in addition to the BCD for a particular codimension $2$ cross-section of $\Sigma$) that uniquely determines a point in phase space $\Gamma.$ 

For the case of 4D Einstein-gravity, one can push $\Sigma$ towards null infinity. In that case, the latter currents reduce to the Bondi-news whereas the BCD is essentially given by the mass-aspect, angular momentum-aspect and additional functions on $S^2$ that provide the Cauchy-data for the solution space (for a review of the solution space in that case see\cite{Barnich:2010eb}). This example is meant to illustrate the way of thinking. As already mentioned, more detailed explanations and examples in simpler settings will be provided somewhere else. 

In summary, so far we have said that the phase space $\Gamma$ can be parametrized by the BCD associated to the symplectic symmetries over a codimension $2$ surface together with their associated symplectic currents. While the BCD will be of our main interest in the following, we want briefly explain that already at this point we are able to learn something. 

\eqref{56} describes the change of the BCD from $B_1$ to $B_2$ caused by the symplectic current passing through a hypersurface $\Sigma$ connecting them. In this way, \eqref{56} reflects a \emph{memory effect.} The independence of the particular choice of $\Sigma$ connecting $B_1$ and $B_2$ in \eqref{56} is a consequence of the \emph{constraint} $d\omega[\delta \Phi,\delta_X \Phi;\Phi]=0.$ In this way, each \emph{symplectic symmetry} $X$ gives rise to a memory effect along an arbitrary hypersurface in $M$ and also to a constraint which altogether reflect the equations of motion. The relation between the concepts symmetry, memory and constraints was recently emphasized in a variety of examples starting with \cite{Strominger:2013jfa,He:2014laa,Kapec:2014opa,Pasterski:2015tva} and references thereof. Here we see, that in the covariant phase space language the equivalence between these concepts becomes obvious and is just reflecting the equations of motion. 

In the remaining part of the chapter, we will explain that the particular way to parametrize the phase space $\Gamma$ can actually be indeed useful to approach various problems. 

Choose a particular point $\Phi \in \Gamma$ by specifying its coordinates, i.e. the BCD $H_X[\Phi]=H_X^{(B_1)}[\Phi]$ for the symplectic symmetries $X$ and their associated symplectic currents. Now, take the latter fixed and vary the BCD. This spans an entire subspace $S \subseteq \Gamma$ on which the BCD then can be seen as coordinates. Thus, $S$ is a submanifold in the phase space $\Gamma.$ However, $S$ has an additional structure. The Poisson-bracket algebra of the generators $H_X=H_X^{(B_1)}[\Phi]$ forms a representation of the Lie-bracket algebra of symplectic symmetries up to central extension. That means, for symplectic symmetries $X,Y$ one has

\begin{equation}
\left\{ H_X,H_Y \right\} = H_{\left[X,Y \right]} + K_{X,Y}
\label{57}
\end{equation}

for c-numbers $K_{X,Y}=K_{X,Y}^{(B_1)}.$ Therefore, the submanifold $S$ is a symplectic manifold on its own. Its coordinates are given by the BCD $H_X$ and their Poisson-bracket algebra is given by \eqref{57}. The part $S$ in phase space $\Gamma$ can therefore be described by a theory on its own right, a ``holographic dual'' associated to the chosen codimension $2$ surface $B_1 \subseteq M.$ 

To summarize, we see that to a chosen codimension $2$ surface $B_1 \subseteq M,$ a holographic dual theory describing a suited part $S \subseteq \Gamma$ can be associated. Choosing a different surface $B_2 \subseteq M$ or different gauge will in general affect the form of \eqref{57} describing the same part $S \subseteq \Gamma.$ Choosing a different $B_2 \subseteq M$ can also lead to a different submanifold in phase space. 

The hope is that the construction of these submanifolds is useful to approach some problems. Usually, the above constructed submanifold $S \subseteq \Gamma$ is too large. However, subalgebras of the algebra of symplectic symmetries will due to \eqref{57} lead to lower-dimensional submanifolds $S'$ in $S.$ Choosing this $S'$ small enough, one is left with a theory that covers a small part of the phase space that might be of interest for a particular problem under consideration. 

Can this be useful?

\subsection{A Microscopic Theory for the Schwarzschild Black Hole}
\label{Kapitel 5.2}

To apply the ideas of the last section \ref{Kapitel 5.1} to a Schwarzschild black hole in Einstein-gravity was essentially the content of \cite{Averin:2018owq}. We recap very briefly the steps. Working in Bondi-gauge, the Schwarzschild-metric fixes a particular point $g_{ab} \in \Gamma$ in covariant phase space. The goal is to find the part of phase space that is responsible for the microstates. The hope is, that the submanifolds constructed in the last section are candidates for this. For the codimension $2$ surface $B_1,$ it is natural to take a cross-section of the event horizon in the hope that the algebra \eqref{57} will get especially simple. 

The next step is then to study closed algebras of symplectic symmetries and their associated submanifolds in $\Gamma.$ For the Schwarzschild black hole $g_{ab},$ there is a simplest choice to start with. Due to the black hole uniqueness theorems, one expects microstate excitations to have the form of residual gauge transformations $\delta g_{ab}=\mathcal L_\xi g_{ab} \in T_{g_{ab}} \Gamma$ for suited vectorfields $\xi.$ Therefore, one is interested in symplectic symmetries $X$ which at $g_{ab} \in \Gamma$ take the form of a residual gauge transformation $X|_{g_{ab}}=\delta_\xi \in T_{g_{ab}} \Gamma.$ Symplectic symmetries of such form and their associated BCD $H_X$ were called gauge aspects in \cite{Averin:2018owq}. \emph{Under the assumption,} that the symplectic symmetries due to the gauge aspects cover enough of the phase space relevant for the Schwarzschild black hole microstates, the BCD parametrizing this submanifold $S \subseteq \Gamma$ as well as its algebra \eqref{57} was determined in \cite{Averin:2018owq}. As explained in the last section, this symplectic submanifold $S \subseteq \Gamma$ provides a theory in its own right and is a candidate for the holographic dual theory of the Schwarzschild black hole. Since this procedure determines the BCD, one is able to infer the form of the residual gauge transformations at $g_{ab}$ which are the candidates for the black hole microstates. 

\subsection{Counting Degrees of Freedom}
\label{Kapitel 5.3}

In the last section, we have explained the construction of a symplectic submanifold $S \subseteq \Gamma$ that is a candidate for the part of the phase space relevant for the microstates of a Schwarzschild black hole. Its coordinates given by the BCD provide observables with the Poisson-bracket algebra \eqref{57}. In this way, we have an explicit theory that provides a candidate for the dual theory governing the Schwarzschild black hole. How can we check whether our candidate theory is correct? 

The first check would be to see whether one can deduce the correct black hole entropy from $S \subseteq \Gamma.$ As explained in \cite{Averin:2018owq}, there are arguments from several directions indicating that the part of phase space responsible for black hole microstates should possess a 2D local conformal symmetry. Therefore, one is tempted to ask whether $S \subseteq \Gamma$ is compatible with this conformal invariance. If so, the observables of $S$ must give rise to a 2D stress-tensor such that its Virasoro-generators fulfill a Virasoro-algebra. Since we know the algebra of observables \eqref{57}, we can search for a Sugawara-construction of these Virasoro-generators out of the BCD over $S.$ This is precisely Carlip's approach to entropy counting in disguise as we will explain in the following. 

In \cite{Averin:2018owq} a projection operator $T_{g_{ab}} \overline{\mathcal{F}} \to T_{g_{ab}} S$ was given, that maps an arbitrary (possibly not gauge-fixed) excitation of a Schwarzschild black hole $g_{ab}$ onto the relevant microstate excitation of $g_{ab}.$ In this way, an arbitrary gauge-excitation $\mathcal L_\xi g_{ab}$ (that in general also contains components that are not tangential to $S$ at $g_{ab}$) is mapped to the relevant symplectic symmetry $X$ tangential to $S.$ It is this mapping $\xi \mapsto X$ from spacetime diffeomorphisms to the vectorfields over $S$ that makes the connection with Carlip's approach clear. In the above mentioned Sugawara-construction, we are searching for symplectic symmetries, i.e. vectorfields $X_n, \overline X_n$ over $S$ such that their generators $H_{X_n}$ and $H_{\overline X_n}$ satisfy via \eqref{57} a $Vir \oplus \overline{Vir}$-algebra. Instead, we can look for diffeomorphisms $\xi_n$ and $\overline \xi_n$ giving rise to a $Vir \oplus \overline{Vir}$-algebra with respect to the spacetime Lie-bracket. This is precisely what we did in chapter \ref{Kapitel 2}-\ref{Kapitel 4}. We furthermore inspected the algebra of the Hamiltonian generators of $\xi_n$ and $\overline \xi_n$ to see that the emerged Virasoro-algebra gives indeed rise to the expected entropy. However, in chapters \ref{Kapitel 2}-\ref{Kapitel 4} we did not employ the mapping $\xi \mapsto X.$ Therefore, so far we only know that the Hamiltonian generators of the $Vir \oplus \overline{Vir}$-diffeomorphisms provide candidates for the Virasoro-generators of a possible would-be CFT governing the black hole microstates. The approach is not sensitive to the details of what this CFT might be.

As already proposed in \cite{Averin:2018owq}, the situation is different once we have figured out our candidate theory $S \subseteq \Gamma.$ We can use the projection operator $\xi_n \mapsto X_n$ and $\overline \xi_n \mapsto \overline X_n$ to obtain with $H_{X_n}$ and $H_{\overline X_n}$ candidates for the Virasoro-generators in $S.$ Precisely this step is sensitive to the choice of $S.$ That means, $X_n$ and $\overline X_n$ and their generators would change if the space $S$ were different. Since we project on $S,$ we are directly probing the degrees of freedom covered by $S.$ Up from here, we can proceed the same way as in the indirect approach. Inspecting the Virasoro-algebra formed by $H_{X_n}$ and $H_{\overline X_n}$ via \eqref{57}, we can count the degeneracy of states and compare it to the expected entropy. If the result were to agree, this would provide substantial consistency check that the theory given by $S \subseteq \Gamma$ is correct and covers all degrees of freedom of the Schwarzschild black hole. Furthermore, it would support that $S$ is indeed conformally invariant thus providing a concrete realization of a Schwarzschild/CFT-correspondence. In case that disagreement is found, one has to enlarge $S$ successively by allowing larger algebras of symplectic symmetries in its construction, up until the procedure is going to converge.

To summarize, with \eqref{30} and \eqref{31} we have given the needed $Vir \oplus \overline{Vir}$-algebra of diffeomorphisms that is needed in the above procedure of projecting directly onto black hole degrees of freedom and counting entropy. These vectorfields were already given in \cite{Averin:2018owq}. Here, we have given their systematic construction. Furthermore, we have provided arguments what singles out the presented $Vir \oplus \overline{Vir}$-diffeomorphisms. Most importantly, we have seen that inspection of the Hamiltonian generators (without projecting directly onto degrees of freedom), we were able to show that the Poisson-bracket algebra consistently leads to the expected Bekenstein-Hawking entropy. Therefore, the $Vir \oplus \overline{Vir}$-diffeomorphisms seem to be the right candidates for the approach described in \cite{Averin:2018owq} and reviewed here. Performing this approach, we leave for future investigations. The purpose of this work here was to provide convincing arguments that the $Vir \oplus \overline{Vir}$-vectorfields are the appropriate diffeomorphisms to use.

\section*{Acknowledgements}
We thank Alexander Gußmann for many discussions on this and other topics in physics and proofreading the work. We thank Malcolm Perry for comments on a preview of this work. We further thank Gia Dvali and Dieter Lüst for discussion during a presentation of the work.

\end{document}